\documentclass[%
 preprint,
 amsmath,amssymb,
 aps,
]{revtex4-2}

\usepackage{graphicx}
\usepackage{bm}
\usepackage{caption}
\usepackage{amsmath}
\usepackage{makecell}
\usepackage{booktabs}
\usepackage{multirow}
\usepackage{array}
\usepackage{xcolor}
\usepackage{setspace}

\begin{document}
\title{Global evidence for a consistent spatial footprint of \\ intra-urban centers}

\author{Shuai Pang$^{1}$}
\author{Junlong Zhang$^{1}$}
\author{Yu Liu$^{1,2}$}
  \email{Corresponding author: liuyu@urban.pku.edu.cn}
\author{Lei Dong$^{1,2}$}
  \email{Corresponding author: leidong@pku.edu.cn}

\affiliation{%
 1. Institute of Remote Sensing and Geographical Information Systems, School of Earth and Space Sciences, Peking University, Beijing 100871, China\\
 2. Ordos Research Institute of Energy, Peking University, Ordos 017000, China
}%

\date{\today}

\begin{abstract}
Urban space is highly heterogeneous, with economic and population activities concentrating in localized centers. However, the global organization of such intra-urban centers remains poorly understood due to the lack of consistent, comparable data. Here we develop a scalable geospatial framework using nighttime light observations to identify over 15,000 intra-urban centers worldwide. We uncover a robust regularity: despite differences in city size, geography, and development context, total urban area scales linearly with the number of centers, implying a roughly constant spatial footprint per center. This macroscopic regularity is underpinned by two independent sublinear scaling laws---center number and urban area both scale with population at closely matched rates---whose ratio cancels to produce the observed linear relationship. At the within-city level, this constancy manifests as a characteristic Voronoi coverage area per center that is consistent across regions, and centers are more regularly spaced than spatial null models predict. As a consequence, polycentric cities maintain stable accessibility as they expand. These findings provide a new empirical foundation for understanding the spatial organization of urban growth.

\end{abstract}

\maketitle

\newpage

\section*{Introduction}
Cities, despite occupying a small fraction of the Earth's terrestrial surface \cite{liu2020high}, function as the primary engines of global population concentration, economic output, and innovation \cite{bettencourt2007growth,glaeser2012triumph,seto2014human,fujita1996economics,glaeser2011cities}. Within these urban agglomerations, socioeconomic activities are far from uniformly distributed; instead, they exhibit a highly heterogeneous spatial structure, gravitating toward localized centers \cite{bertaud2018order}. Identifying these internal centers is fundamental to understanding the spatial logic of cities, as they govern population distribution \cite{clark1951urban,anas1998urban}, rent gradients \cite{liotta2022testing}, housing prices \cite{zheng2008land}, traffic flows \cite{louf2013modeling}, and service accessibility \cite{weiss2018global,cattaneo2021global,cattaneo2024worldwide}. Beyond their functional roles, how these centers are arranged in space remains an open question. Classical central place theory envisions urban systems organizing into an orderly partitioning of space, where centers serve discrete hinterlands to minimize travel costs and maximize service efficiency \cite{christaller1933zentralen,berry1958note}. Yet, global empirical evidence for such patterns has remained elusive.

This knowledge gap stems largely from the lack of a consistent and scalable methodology to delineate intra-urban centers globally. The canonical approach identifies (employment) subcenters from census data, economic surveys, and predefined statistical units (e.g., census tracts) \cite{griffith1981evaluating,mcdonald1987identification,mcmillen2001nonparametric,redfearn2007topography}, but these data are available only for a limited number of countries with detailed economic censuses. Mobile phone and social media data can partially compensate for this lack of coverage \cite{cai2017using,yan2023new}, yet their availability remains limited, hindering global comparability. Gridded population data, while globally available, primarily reflect residential distribution rather than economic activity concentration, which can diverge substantially in cities where employment centers are spatially separated from residential areas (Supplementary Fig.~1 and Supplementary Text). Moreover, efforts to identify urban centers based on population distribution have largely been confined to specific countries or regions \cite{liu2016polycentric,li2018did}, with few mapping exercises conducted at the global scale. Recently, the European Union's Urban Centre Database based on Global Human Settlement Layer (GHSL) \cite{rivero2024ghs, melchiorri2024multi} represents important progress in mapping urban cores worldwide, yet it delineates entire agglomerations as single units rather than identifying the multiple centers nested within them. Consequently, the polycentric structure of modern metropolitan growth remains poorly captured by existing representations.

To address these limitations, we develop a geospatial framework that leverages nighttime light data to construct a harmonized global map of intra-urban centers. Nighttime light is a well-established proxy for economic activity \cite{donaldson2016view,burke2021using, chen2011using, henderson2012measuring, henderson2018global,chen2022global}, and has been increasingly used for urban structure identification \cite{chen2017new, yang2021using, jie2024identifying}. Building on this foundation, we treat light intensity as a continuous surface of urban activity. In this study, a \textit{city} refers to a contiguous urban agglomeration delineated by GHSL, independent of administrative boundaries, ensuring global consistency \cite{dong2024defining}. An \textit{urban center} (or simply \textit{center}) denotes a localized concentration of activity within a city. Analogous to mapping mountainous terrain, we operationally identify centers as local maxima in the nighttime light intensity landscape (Fig. 1). 

Analyzing this global map reveals substantial diversity in the number and arrangement of centers across regions. Beneath this heterogeneity, however, lies a robust empirical regularity: across cities worldwide, the total urban area scales linearly with the number of centers, implying a roughly constant spatial footprint per center. That aggregate constancy is stable under a wide range of methodological parameters and different city boundary delineations, and holds for both 2015 and 2020 datasets. Its micro-level origin lies in the spatial organization within cities: specifically, the spatial catchment area of each individual center exhibits a characteristic size. These regularities imply that as cities grow, they tend to accommodate expansion through the multiplication of centers rather than the expansion of existing ones.

Our study makes two key contributions to the existing literature. First, we focus on the distribution of activity \textit{within} cities. This within-city perspective reduces systematic differences in nighttime light intensity arising from geography, lighting policies, and other inter-city factors, so that centers can be identified as long as within-city light variation is sufficiently pronounced (Supplementary Fig.~2). It also extends previous discussions on the spatial distribution of cities as points in space \cite{mori2020common} to the intra-urban scale, providing a data foundation for comparing internal urban structure across cities and regions. Second, by introducing center number as a quantitative measure, our study offers a more micro-level perspective on the population–area analyses central to urban scaling theory. Specifically, we show that the linear scaling between urban area and center number can be decomposed into two sublinear scaling relationships: center number scales with population as $N \sim P^{\gamma}$ \cite{louf2013modeling}, and urban area scales with population as $A \sim P^{\alpha}$ \cite{bettencourt2013origins}. Empirically, we find that these two exponents are closely matched ($\alpha \approx \gamma$), so that their ratio cancels to produce the observed linear center–area scaling. Our framework thus bridges the spatial structure of polycentricity with the macro-level regularities of urban scaling.

\section*{Results}
\subsection*{Identification of Intra-Urban Centers}
We outline the core datasets and methodology below; detailed parameters, computation procedures, and additional validation analyses are provided in the Methods section and Supplementary Figs.~2--8.

The primary inputs for identifying intra-urban centers are the Global Human Settlement Layer Urban Centre Database (GHSL-UCDB) and the VIIRS nighttime light data. We extract urban boundaries from the GHSL-UCDB to mask the light data, and then apply a contour-based detection algorithm adapted from the Localized Contour Tree Method~\cite{chen2017new} to the masked data. Nighttime light intensity is treated as a continuous surface reflecting spatial variation in urban activities, and local intensity maxima serve as center candidates. A candidate qualifies as a true center if its enclosed contour area exceeds a minimum threshold. The threshold is set at 4 km$^2$ for the main analyses, corresponding to a spatial footprint of roughly 2 km $\times$ 2 km, which is suitable for capturing localized activity concentrations. We also demonstrate that our main findings are robust across thresholds ranging from 1 to 10 km$^2$, as shown later.

Because no existing dataset offers globally comparable delineations of intra-urban centers, we adopted complementary validation strategies. First, we compiled planning documents from four cities—Seattle, Portland, Beijing, and London—that explicitly designate government-identified urban centers. After georeferencing these planned centers, we found that 30 of the 34 planned centers across the four cities are matched by our detected centers (located within 1 km from the planned centers; Supplementary Fig.~3). Second, we compared our centers with hotspots derived from a recursive head/tail breaks method applied to nighttime light data by Ren et al. \cite{ren2024characterizing, ren2024topology}. Most hotspots are matched by our detected centers (Supplementary Fig.~4). Third, we leveraged the reasoning capabilities of a multimodal large language model to conduct a post-assessment of remote sensing imagery surrounding all detected centers, in order to estimate the proportion of false positives. The prompt specified criteria for common false-positive categories, including chemical plants, highway interchanges, large-scale greenhouses, power plants, and so on. The model was instructed to return the classification results. Overall, approximately 4\% of centers were flagged as likely false positives, predominantly chemical plants; see Supplementary Table 1 and Supplementary Fig.~5 for details.

Together, these independent validation approaches confirm that our detected centers reliably capture localized concentrations of urban activity across diverse geographic contexts.

\subsection*{Center-Area Scaling}

Globally, we identify 15,533 centers across 9,478 cities. Figure~1b presents the spatial distribution of urban centers, reflecting both widespread urbanization and pronounced regional concentration. High densities of centers are observed in large metropolitan regions, such as the eastern United States, central and western Europe, and eastern China. China hosts the largest number of detected centers (2,908), followed by India (2,124) and the United States (1,337), consistent with their large urban populations and extensive built-up areas. Within individual countries, however, centers are distributed unevenly. In China, 92\% of centers are located east of the Heihe-Tengchong Line \cite{hu1983discussions}---an area accounting for roughly one-third of the national territory. In India, approximately 40\% of centers are located within the Indo-Gangetic Plain, which occupies about 20\% of the country's land area. In the United States, 69\% of centers are concentrated in the top five census divisions that together account for only 28.1\% of the national area. This unevenness extends to the city scale, where the number and spatial arrangement of centers vary substantially across urban agglomerations.

At the city scale, Fig.~1c and Supplementary Fig.~6 illustrate the spatial arrangement of detected centers in representative cities, revealing substantial variation in both center multiplicity and spatial dispersion. For example, Los Angeles exhibits a highly dispersed configuration with 92 detected centers, whereas London and Luanda display more compact patterns with 30 and 10 centers, respectively.

Beyond this geographic heterogeneity, a robust quantitative regularity emerges. The average area per center---computed as the total GHSL urban area divided by the number of detected centers---remains approximately constant as the number of centers increases (Fig. 2a). This pattern is captured by a scaling relationship between the number of centers $N_i$ and the urban area $A_i$ of city $i$:

\begin{equation}
    A_i = A_0N_i^\beta
\end{equation}

\noindent where $\beta = 0.98 \pm 0.06$ (Fig. 2b). The near-unity exponent confirms that urban area scales near-linearly with center number, implying that each additional center is associated with a roughly constant increment of urban area.

To test robustness, we systematically varied the minimum contour area threshold from 1 to 10~km$^2$ and repeated the full analysis for both the 2020 and 2015 datasets independently. As expected, the average area per center increases with the threshold in both years (Fig. 2c). However, the scaling exponent $\beta$ remains stable near unity across the full parameter range for both time periods (Fig. 2d), confirming that the linear scaling relationship is independent of parameter choice and temporally stable. This relationship also holds under an alternative city boundary definition (Supplementary Fig.~9). These robustness checks confirm that the linear center–area scaling reflects a structural property of cities rather than an artifact of particular methodological settings.

\subsection*{Decomposition of Scaling Laws}
The linear scaling between urban area and center number is not confined to the global pooled analysis; it also holds at the country level. Figure~3a presents scaling results for 10 large countries, with five highlighted individually. Despite vast differences in geography and urban morphology, the scaling exponents consistently cluster near unity, confirming that this regularity transcends any single region or urban tradition. Supplementary Fig.~11 presents the robustness check by varying the minimum contour area threshold.

The scaling documented above can be situated within a theoretical framework linking urban polycentricity to allometric growth. Building on the Fujita--Ogawa framework \cite{fujita1982multiple}, Louf and Barthelemy~\cite{louf2013modeling} proposed an analytical, out-of-equilibrium model in which the emergence of subcenters is driven by traffic congestion. In their framework, individuals choose workplaces by trading off wage attractiveness against commuting costs. As a city's population grows, congestion at existing centers eventually exceeds a critical threshold, making it advantageous for workers to switch to less congested alternatives---thereby activating new centers. This mechanism yields a quantitative prediction: the number of centers $N$ scales sublinearly with population $P$ as $N \sim P^{\gamma}$, where $\gamma =\mu/(\mu+1)$ and $\mu$ characterizes the sensitivity of travel costs to congestion~\cite{louf2013modeling}. 

To connect this population--center relationship to the center--area scaling documented in this study, a second relationship is required: how urban area scales with population. Urban scaling theory supplies this link, establishing that built-up area grows sublinearly with population, $A \sim P^{\alpha}$ with $\alpha < 1$~\cite{bettencourt2013origins,ribeiro2023mathematical}. This sublinear relationship has been extensively validated across countries and historical periods; Bettencourt's theoretical framework predicts $\alpha = 5/6 \approx 0.83$~\cite{bettencourt2013origins}. Combining these two scaling laws by eliminating $P$ yields $A \sim N^{\,\alpha/\gamma}$; if the two exponents are comparable ($\alpha \approx \gamma$), their ratio approaches unity, recovering the linear scaling $A \sim N^1$ observed in our data.

Figure~3b--e confirms that both sublinear relationships hold empirically. Using US cities as an illustration, the number of centers scales with population as $N \sim P^{\gamma}$ with $\gamma = 0.79 \pm 0.07$ (Fig. 3b), and urban area scales with population as $A \sim P^{\alpha}$ with $\alpha = 0.84 \pm 0.04$ (Fig. 3c). Figure~3d extends this analysis to 10 countries, showing that both the directly estimated center–area exponent $\beta$ (blue) and the exponent predicted by the decomposition $\alpha/\gamma$ (purple) cluster around unity. At the individual-country level, the rate at which centers multiply ($\gamma$) and the rate at which urban area expands ($\alpha$) in response to population growth are closely matched (Fig. 3e). This near-equality of the two growth rates causes their ratio to approach one, producing the linear center–area scaling observed in this study.

This balance between the two exponents is not guaranteed  \emph{a priori}, because the two scaling laws arise from distinct mechanisms: center multiplication is driven by congestion dynamics according to Louf and Barthelemy~\cite{louf2013modeling}, whereas areal expansion reflects the space-filling geometry and infrastructure network properties of growing cities~\cite{bettencourt2013origins}. That these two processes respond to population growth at nearly the same rate implies a tight coupling: as a city grows, the pace at which congestion triggers new centers matches the pace at which the urban footprint expands, so that each new center is effectively paired with a roughly constant increment of land. If center formation outpaced spatial expansion ($\gamma > \alpha$), centers would become increasingly crowded; if it lagged ($\gamma < \alpha$), cities would develop large under-served gaps between centers. The observed balance ($\alpha \approx \gamma$, and hence $\beta \approx 1$) indicates that cities maintain a spatial equilibrium in which the supply of new activity nodes keeps pace with the supply of new urban space.

\subsection*{Spatial Distribution of Centers within Cities}

The city-level scaling relationships described above characterize how urban space is organized across cities. However, such aggregate patterns do not reveal how space is allocated among individual centers within a city. In principle, a stable average area per center could mask substantial internal heterogeneity --- for example, dominant central hubs surrounded by smaller peripheral centers, or compact inner-city centers coexisting with expansive suburban ones. We therefore examine whether the regularity in per-center spatial footprint extends to the within-city scale.

We begin by analyzing the spatial partitioning induced by centers using Voronoi diagrams, which assign each location in a city to its nearest center. Figure~4a illustrates this construction for Los Angeles, where individual Voronoi polygons exhibit comparable areas despite substantial variation in center location and surrounding urban context (mean = 49 km$^2$, SD = 20 km$^2$). To assess whether this apparent regularity generalizes, we analyze Voronoi-based coverage areas across countries and within individual metropolitan areas. Figure~4b compares the distribution of Voronoi areas for centers in cities across China, Germany, the United Kingdom, the United States, and Japan, revealing a consistent mean coverage area---typically between 42 and 47 km$^2$ per center. Figure~4c further demonstrates that similar regularities hold within individual metropolitan areas, where centers in different spatial positions nonetheless exhibit comparable coverage areas.

To ascertain whether the observed regularity reflects a genuine spatial property rather than an artifact of spatial partitioning, we compared the empirical distribution of Voronoi areas against two null models: complete spatial randomness (CSR) and sequential spatial inhibition (SSI) (Fig. 4d). The CSR model places points independently and uniformly within each city, representing a baseline with no spatial structure. The SSI model enforces a minimum separation distance between points, with each point modeled as a hard disk whose area equals the minimum contour area threshold (4 km$^2$). This second model accounts for the spatial exclusion inherent in our detection method --- since two centers cannot be identified within overlapping contour areas --- and thus provides a more stringent test of spatial regularity. To avoid edge effects, all spatial metrics are computed only for internal centers whose Voronoi cells do not intersect the city boundary.

The empirical Voronoi polygons exhibit a lower coefficient of variation than both null models: $CV_{\mathrm{data}} \approx 0.37$, compared with $CV_{\mathrm{CSR}} \approx 0.58 \pm 0.04$ for CSR and $CV_{\mathrm{SSI}} \approx 0.47 \pm 0.04$ for SSI ($p < 0.01$; Fig.~4e). Consistent with this, empirical centers are more widely spaced, with a mean nearest-neighbor distance of $\overline{d}_{\mathrm{data}} \approx 4.51$~km, significantly larger than under CSR ($\overline{d}_{\mathrm{CSR}} \approx 3.06 \pm 0.12$~km) or SSI ($\overline{d}_{\mathrm{SSI}} \approx 3.87 \pm 0.11$~km, both $p < 0.01$; Fig.~4e). The improvement from CSR to SSI confirms that part of the observed regularity is attributable to the minimum separation enforced by our detection algorithm. Crucially, however, the empirical pattern remains significantly more regular than SSI, indicating that actual intra-urban centers are more evenly spaced than can be explained by the exclusion constraint alone. Whereas randomly placed points, with or without a hard-core repulsion, frequently leave large spatial gaps or form local clusters, actual centers maintain greater and more uniform spatial separation, producing a more orderly tiling of the urban fabric.

\subsection*{Polycentric Configuration and Accessibility}

The characteristic coverage area of centers has direct implications for urban accessibility. We quantify this by comparing the population-weighted average distance to the nearest center under two scenarios: the observed configuration, and a counterfactual baseline in which all activity is consolidated into a single primary core (Fig. 5a).

As shown in Fig. 5b, the two scenarios exhibit divergent accessibility trajectories as cities grow. Under the monocentric assumption, the average distance from the population to the main center increases monotonically as a power function. In sharp contrast, the empirical population-weighted distance to the nearest center remains approximately constant, fluctuating around a stable baseline regardless of whether a city possesses two or twenty centers. The result is robust to alternative population datasets and to variations in center detection parameters (Supplementary Fig.~12), confirming that it reflects a geometric property of polycentric organization rather than a data-specific artifact.

We note that this constancy is not a geometric tautology. While adding any service point trivially reduces the average distance to the nearest point, the empirical finding is more specific: the population-weighted mean distance remains approximately \textit{constant} across cities spanning a wide range of sizes and center counts. This constancy requires that new centers emerge at locations that maintain a roughly stable service radius per center. If, for example, new centers clustered near existing ones or emerged preferentially at the urban periphery, the population-weighted nearest-center distance would vary systematically with city size. The observed invariance thus reflects the specific spatial arrangement of centers---consistent with the linear scaling ($\beta \approx 1$) documented above---rather than a necessary consequence of polycentricity per se.

This stability indicates that the emergence of new activity centers serves as a spatial adaptation to growth. By maintaining a roughly constant spatial footprint per center, cities effectively ``tile" the landscape with accessible hubs, ensuring that the average resident's proximity to economic activity is decoupled from the overall size of the metropolitan area.

\section*{Discussion}

By constructing a global dataset of intra-urban centers, this study reveals a robust empirical regularity beneath the diversity of urban spatial organization: each urban center is associated with a roughly constant spatial footprint. This regularity is not an artifact of a particular operational definition---it persists across detection thresholds, time periods, and city boundary delineations---suggesting that polycentric configurations emerge as a generic spatial response to urban growth rather than as idiosyncratic outcomes of local planning.

Our findings resonate with Central Place Theory (CPT) \cite{christaller1933zentralen}, which assumes that centers of the same level serve equal-sized hinterlands in a regular spatial arrangement, while higher-order centers command progressively larger catchment areas. The observed regularities, a characteristic Voronoi area per center, parallel CPT's prediction of evenly partitioned market areas among same-level centers. Additionally, in our framework, the minimum contour area threshold functions as a filter: a low threshold retains centers across all levels, whereas a high threshold selects only higher-order centers. As the threshold increases, the average area per center grows (Fig. 2c), consistent with the expectation that higher-order centers serve larger hinterlands. While our data do not test CPT's specific predictions regarding nesting ratios or hexagonal geometries, the qualitative alignment suggests that the observed regularities may reflect the spatial logic broadly consistent with central place principles.

The decomposition analysis further reveals that the linear center–area scaling is not an isolated empirical observation but the emergent outcome of two independent sublinear relationships: center number and urban area both scale with population at comparable rates. This near-equality is non-trivial---the two processes are governed by distinct mechanisms (congestion-driven center formation versus infrastructure-mediated spatial expansion), yet they respond to population growth in lockstep across countries and continents. Had center formation outpaced spatial expansion, cities would exhibit increasingly dense packing of centers; had it lagged, growing gaps between centers would erode accessibility. The observed balance suggests that urban systems self-organize toward a spatial equilibrium in which new activity nodes and new urban space are supplied at commensurate rates.

Despite these advances, several limitations remain. A key constraint lies in using nighttime light intensity as a proxy for urban activity, which may not fully capture urban vibrancy across all contexts \cite{bluhm2022top,wu2023measuring}. Geographic and cultural biases in nighttime light, such as energy-saving policies, irregular electricity supply, or weak nighttime illumination, may lead to under-identification in certain regions, while brightly lit non-urban features may occasionally be misclassified as centers. Detailed analyses of these limitations are presented in Supplementary Figs. 5, 7, and 8. 

Future research should explore the linkages between the intra-urban centers and human mobility behavior \cite{louail2014mobile,mazzoli2019field,bassolas2019hierarchical,dong2022universality,schlapfer2021universal}. These mobility behavioral constraints offer a possible interpretation for why centers across diverse contexts are associated with similar spatial extents: individuals' finite willingness to travel may impose a characteristic radius on the catchment area that each center can effectively serve.

\section*{Methods}

\subsection*{Data}

We integrated five datasets: Global Human Settlement Layer -- Urban Centre Database (GHSL-UCDB) \cite{melchiorri2024multi}, nighttime light data from the Visible Infrared Imaging Radiometer Suite (NPP-VIIRS) \cite{elvidge2021annual}, Foursquare Open Source Places (POI, \url{https://opensource.foursquare.com/os-places}), GeoNames geographical database (\url{https://www.geonames.org}), and WorldPop Population Counts \cite{bondarenko2025population}. GHSL-UCDB provides the spatial extent of cities; NPP-VIIRS provides a proxy for aggregated human activity used to identify intra-urban centers; Foursquare Open Source Places support center labeling and functional characterization; GeoNames supplies standardized place names; and WorldPop provides population estimates for scaling regressions and accessibility analyses.

GHSL-UCDB delineates global settlements based on population size, population density, and built-up area, following the Degree of Urbanisation framework \cite{pesaresi2024advances}. We used the 2015 and 2020 high-density cluster, defined as contiguous 1 km$^2$ grid cells with at least 1,500 people per km$^2$ or over 50\% built-up area, and a minimum population of 50,000.

The VIIRS Day/Night Band measures radiance in units of $\text{nW}\cdot\text{cm}^{-2}\cdot\text{sr}^{-1}$ with a spatial resolution of $\sim$ 500 m. We used the 2015 and 2020 annual composites (version ``vcmslcfg''), which filter out transient light sources such as sunlight, moonlight, biomass burning, and aurora. Background noise was removed by setting low-intensity values to zero using a standard brightness threshold \cite{elvidge2021annual}.

Foursquare Open Source Places provides points of interest across 1,245 categories, including business, government, retail, dining, and transportation. GeoNames integrates multi-source geographic information, including place names and administrative units. WorldPop offers population mosaics at 1-km resolution derived from top-down models calibrated to national census data; we used the unconstrained product for population-based analyses.

All datasets were referenced to the WGS84 geographic coordinate system. Urban areas were calculated geodesically, and population and brightness were aggregated within GHSL-UCDB boundaries using spatial masking.

\subsection*{Contour Methods to Identify Local Activity Centers}
We identified intra-urban activity centers using a modified Localized Contour Tree method \cite{chen2017new}. In this framework, nighttime light intensity is conceptualized as a continuous surface of aggregated urban activity, with local maxima representing candidate centers. Because contour extraction requires spatially smoothed input data, we followed established practice and applied a Gaussian filter (3 $\times$ 3 pixels, $\sigma$= 5) to the nighttime light imagery to enhance spatial continuity and eliminate ill-formed contours \cite{chen2017new}. Contours were extracted at 1 $\text{nW}\cdot\text{cm}^{-2}\cdot\text{sr}^{-1}$ intervals within each GHSL-defined urban boundary. Higher-intensity contours enclose progressively smaller areas, analogous to elevation contours in topographic analysis.

To distinguish meaningful centers, we introduced a minimum contour area threshold---the main tuning parameter of the approach---which defines the smallest spatial extent considered representative of a localized activity concentration. As illustrated in Fig. 1a, a contour is considered a valid peak contour (seed contour) if it satisfies two conditions: (i) it does not contain any smaller contour that also meets the area criterion (i.e., no nested qualified contour exists within it), and (ii) its light value is higher than that of surrounding contours, ensuring it represents a local maximum rather than a saddle or valley. For each retained contour, the brightest pixel is selected as the representative center location. Robustness checks of the main findings across different parameters are shown in the Supplementary Figs. 9, 11, and 12.



Because detection relies on nighttime light intensity within GHSL-defined urban extents, not all urban areas yield detectable centers. In total, 1,873 cities (16.5\%, Supplementary Fig.~7) exhibit no detectable centers. These cases arise for two primary reasons. First, some densely populated settlements---mainly in Nigeria, China and India---meet GHSL criteria for population density and built-up area but display diffuse nighttime light patterns without pronounced local maxima. In such cases, residential clustering is evident, yet neither nighttime light nor high-resolution daytime imagery reveal localized concentrations of activity (Supplementary Fig.~8a--c). Second, cities in low-illumination contexts, such as Pyongsong (North Korea), exhibit clearly defined urban cores in satellite imagery but insufficient nocturnal lighting for reliable detection (Supplementary Fig.~8d). These limitations motivate future integration of complementary data sources, such as multispectral imagery, to improve coverage in low-light environments.

\subsection*{Attribute Enhancement}
Beyond center identification, we enriched the dataset by labeling main centers, assigning place names, and characterizing functional types. First, we classified main centers and subcenters. In monocentric cities, the single detected center was designated as the main center. In polycentric cities, we calculated POI density within each center’s seed contour; the center with the highest POI density was labeled the main center, and all others were classified as subcenters. Second, each center was assigned a place name and a functional category. Place names were obtained by matching centers to the nearest GeoNames record within a 1-km buffer. Functional classification was based on POIs aggregated within the same buffer and grouped into six major categories: Mixed, Consumption, Business, Public Service, Transport, and Industrial. We calculated the Shannon entropy $H = -\sum_{i=1}^{5} p_i \log p_i$ over the five non-Mixed categories to quantify functional diversity. Centers with $H > 1.5$ were labeled as Mixed; otherwise, centers were assigned the dominant POI category. Overall, mixed-use centers constitute the majority, accounting for nearly 60\% of all detected centers.

Because centers represent spatial extents rather than point locations, we provide the full contour geometries alongside point coordinates in the public dataset to support future analyses.

\clearpage

\begin{figure}[h]
    \centering
    \includegraphics[width=\textwidth]{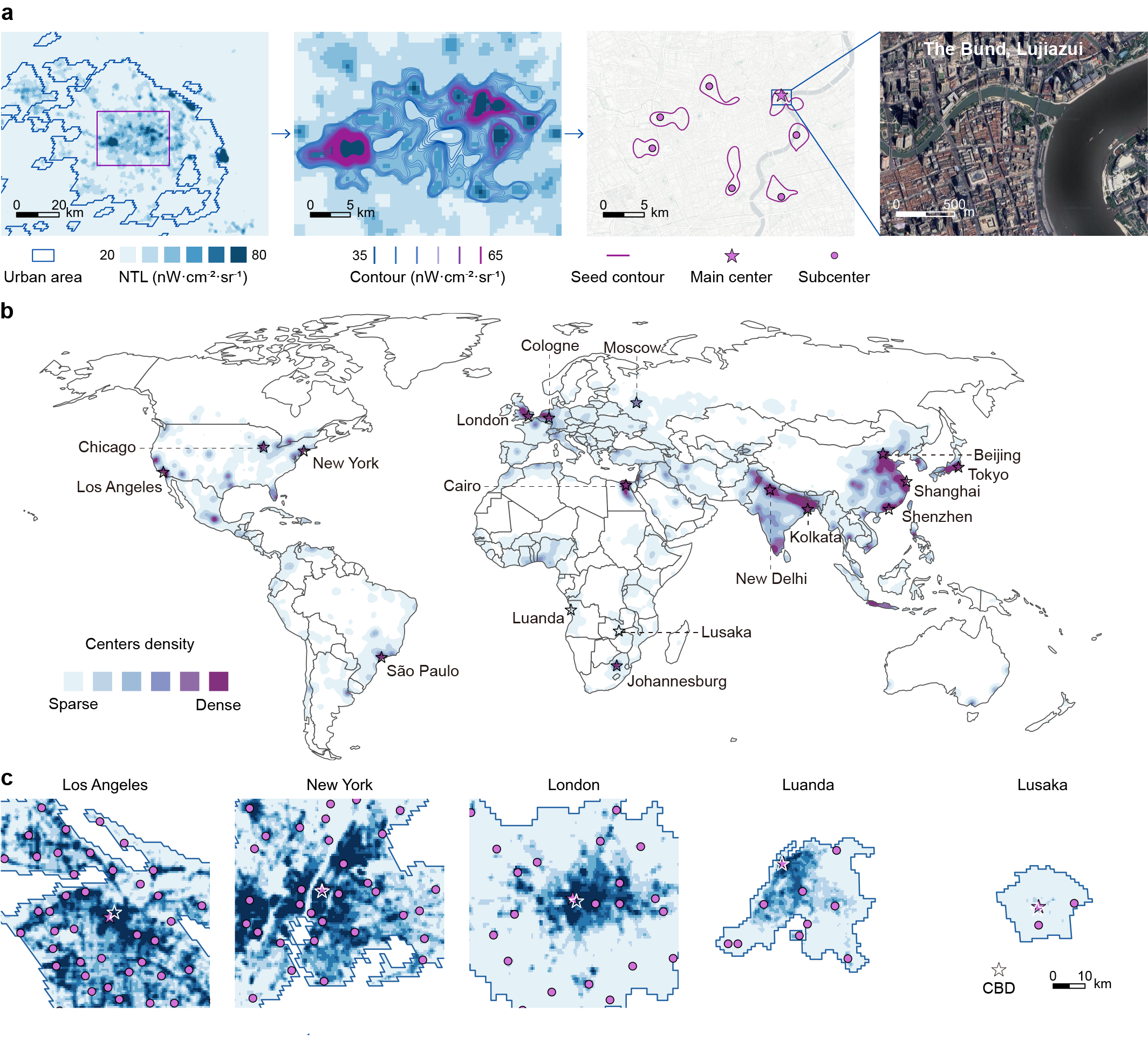}
    \captionsetup{font={footnotesize, stretch=0.8}}
    \caption{\textbf{Urban center identification.} (\textbf{a}) The identification workflow using Shanghai, China, as an example. From left to right: nighttime light distribution; contour maps of the region; seed contours and identified centers (under the baseline minimum contour area of 4 km$^2$); and a close-up satellite view of the identified main center (Imagery \copyright 2026 Airbus, CNES / Airbus, Maxar Technologies, Map data \copyright 2026 Google). Center with the highest POI density is labeled as the main center, while the rest are defined as subcenters. (\textbf{b}) Heatmap of centers worldwide. Developed or densely populated regions typically have a high density of centers. (\textbf{c}) Urban centers in Los Angeles, US; New York, US; London, UK; Luanda, Angola; and Lusaka, Zambia. The locations of the CBDs are collected from official government reports and online documents. The spatial proximity of the identified main centers (purple stars) to the CBDs (white stars) partially validates the accuracy of our center identification procedure.}
\label{Methods}
\end{figure} 
\clearpage

\begin{figure}[h]
    \centering
    \includegraphics[width=15 cm]{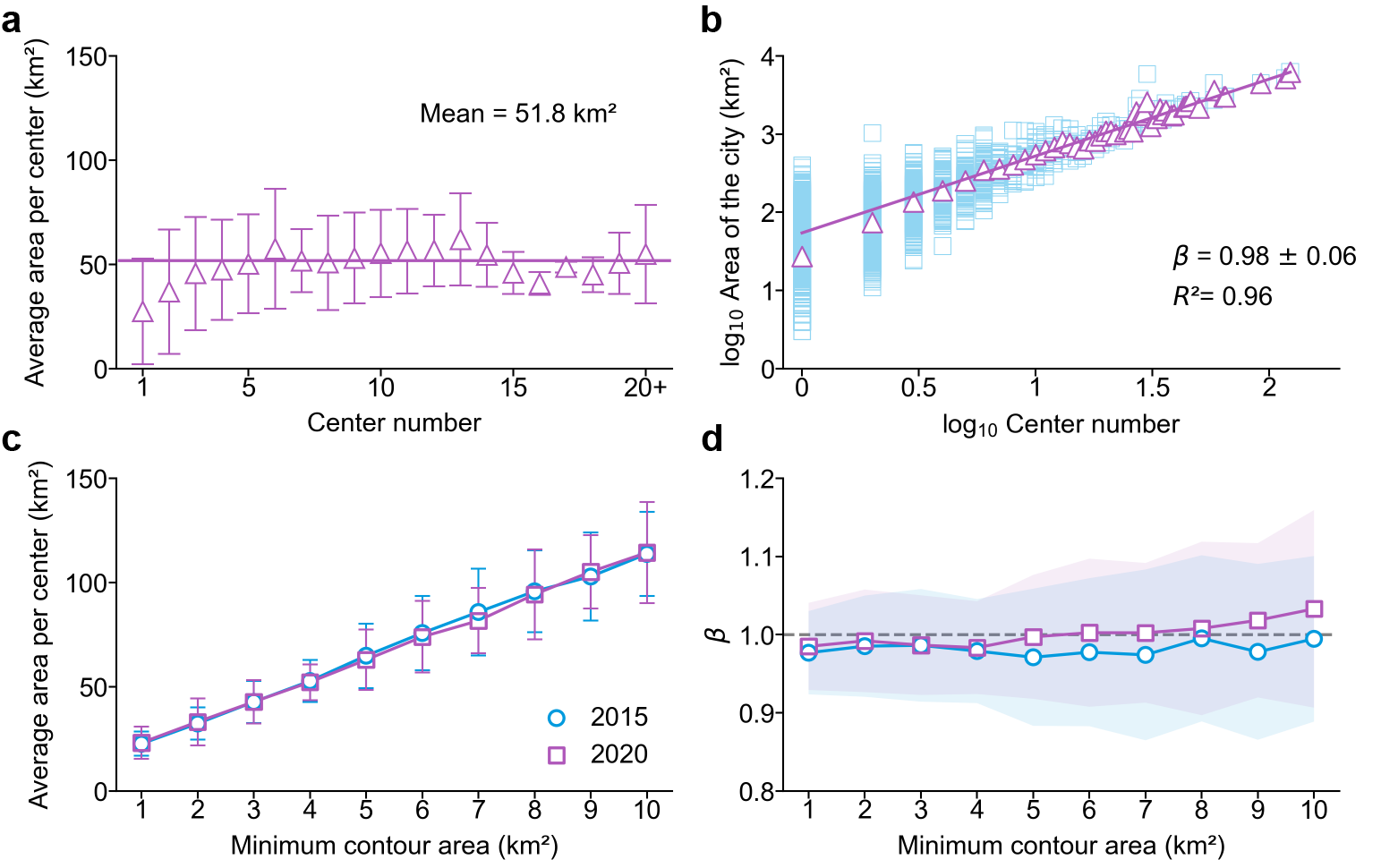}
    \captionsetup{font={footnotesize, stretch=0.8}}
    \caption{\textbf{Linear scaling relationship between city area and center number.} (\textbf{a}) Average area per center as a function of the center number of cities (2020, minimum contour area = 4 km$^2$). Cities with $\ge$ 20 centers are aggregated. Triangles indicate the mean, error bars represent the standard deviation and the horizontal line denotes the average value of the means for cities with at least 4 centers. (\textbf{b}) Scaling analysis between the total built-up area and center number (2020, minimum contour area = 4 km$^2$). Squares represent individual cities, and triangles indicate group means. The line represents the fitting to the group means (triangles) of cities with at least 4 centers on a log-log scale. Sensitivity of $\beta$ to the minimum number of centers included in the regression is shown in Supplementary Fig.~10. (\textbf{c}) Average area per center versus the minimum contour area threshold for 2015 and 2020. (\textbf{d}) Scaling exponent between the total built-up area and center number versus the minimum contour area threshold for 2015 and 2020. The exponent $\beta$ remains close to unity across all settings.}
\label{Overview}
\end{figure} 
\clearpage

\begin{figure}[h]
    \centering
    \includegraphics[width=16 cm]{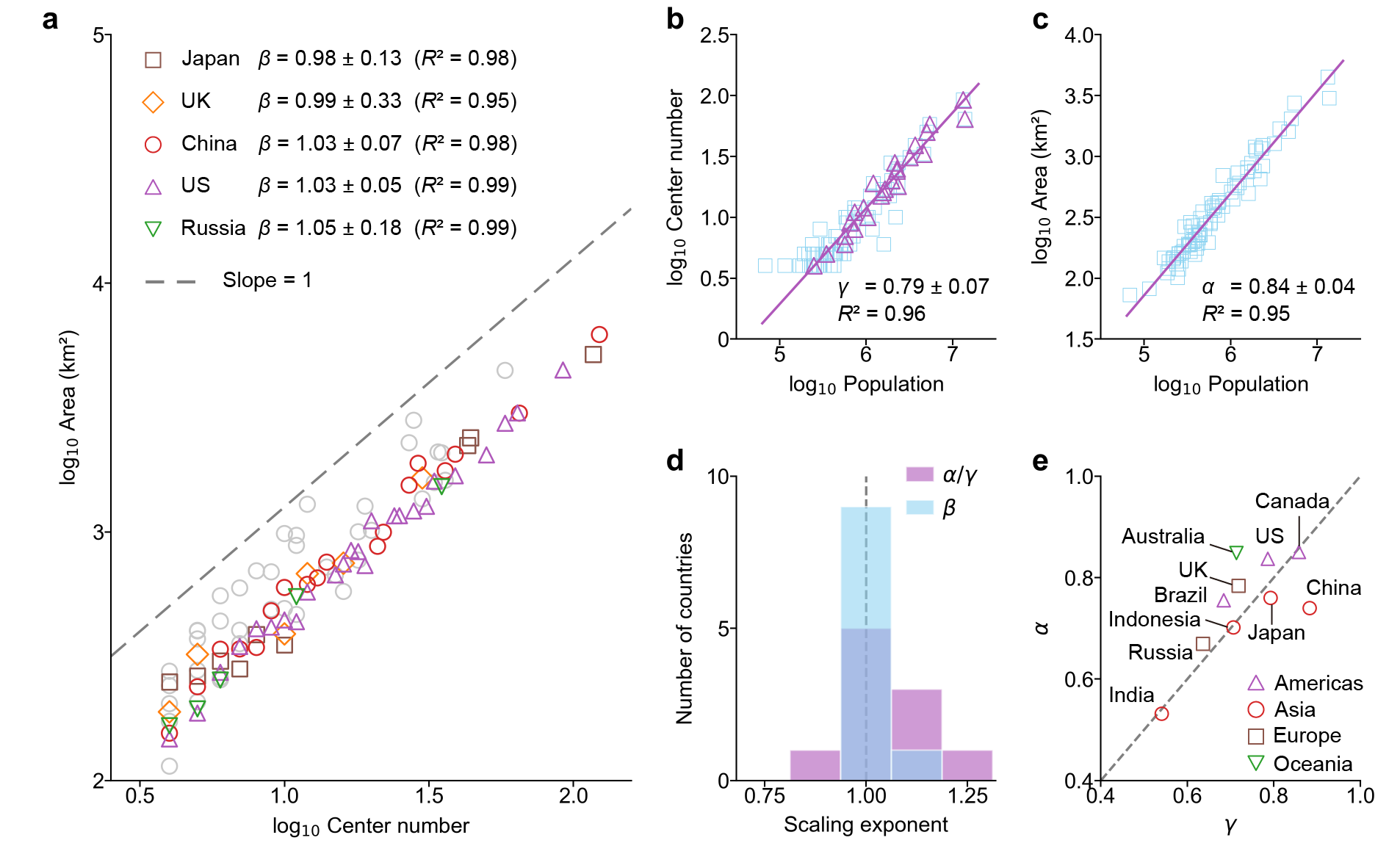}
    \captionsetup{font={footnotesize, stretch=0.8}}
    \caption{\textbf{Consistency between center--population and area--population scaling laws.} (a) Area--center scaling relationship ($A\sim N^\beta$) for cities with $\geq$ 4 centers across 10 large countries. Group means are shown for 5 highlighted countries; group means of the remaining 5 countries are plotted as light-gray open circles. (b) Center--population scaling ($N \sim P^\gamma$) for US cities with $\geq$ 4 centers. (c) Area--population scaling ($A \sim P^\alpha$) for US cities with $\geq$ 4 centers. (d) Distribution of the direct scaling exponent $\beta$ (blue) and the ratio $\alpha / \gamma$ (purple) across 10 countries. (e) Cross-country comparison of $\alpha$ against $\gamma$, colored by continent. Scaling regressions are fitted to group means (area--center and population--center) or individual cities (area--population) on a log-log scale. Results are based on data from 2020 using a minimum contour area threshold of 4~km$^2$.}
    \label{Chain}
\end{figure}
\clearpage

\begin{figure}[h]
    \centering
    \includegraphics[width=\linewidth]{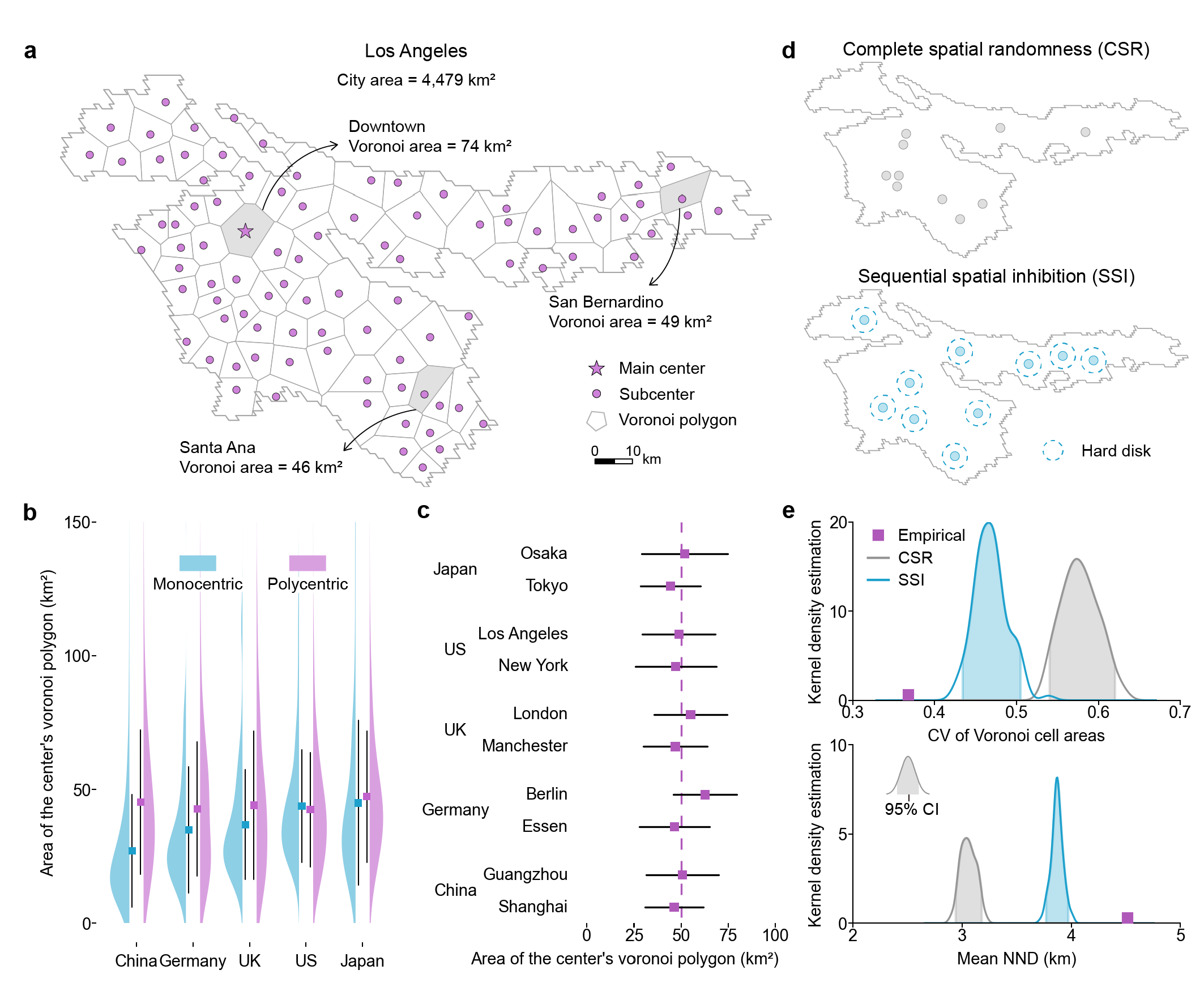}
    \captionsetup{font={footnotesize, stretch=0.8}}
    \caption{\textbf{Spatial distribution of centers within the city.} (\textbf{a}) Voronoi diagram of centers in Los Angeles, US. (\textbf{b}) Area distribution of centers' Voronoi polygons in monocentric and polycentric cities for China, Germany, the UK, the US and Japan. The average coverage area of centers in polycentric cities remains stable across different countries. (\textbf{c}) Area distribution of centers' Voronoi polygons in ten cities. Squares indicate the mean, error bars represent the standard deviation and the dashed purple line denotes the average value of the means. (\textbf{d}) Schematic of the null models of complete spatial randomness (CSR) and sequential spatial inhibition (SSI). The sampled points are independent in CSR, but are constrained by a minimum repulsion distance in SSI. (\textbf{e}) Comparison between the empirical distribution of centers and null models, measured by the coefficient of variation (CV) of Voronoi areas (top) and mean nearest-neighbor distance (NND, bottom). The kernel density estimates show 100 Monte Carlo simulations. The empirical CV (0.37) is significantly lower than both the CSR (0.58 $\pm$ 0.04) and SSI (0.47 $\pm$ 0.04, both $p < 0.01$), and the empirical mean NND (4.51 km) significantly exceeds the CSR (3.06 $\pm$ 0.12 km) and SSI (3.87 $\pm$ 0.11 km, both $p < 0.01$), indicating that real centers are more regularly spaced than expected under either random baseline.}
\label{Spatial}
\end{figure}

\begin{figure}[h]
    \centering
    \includegraphics[width=\textwidth]{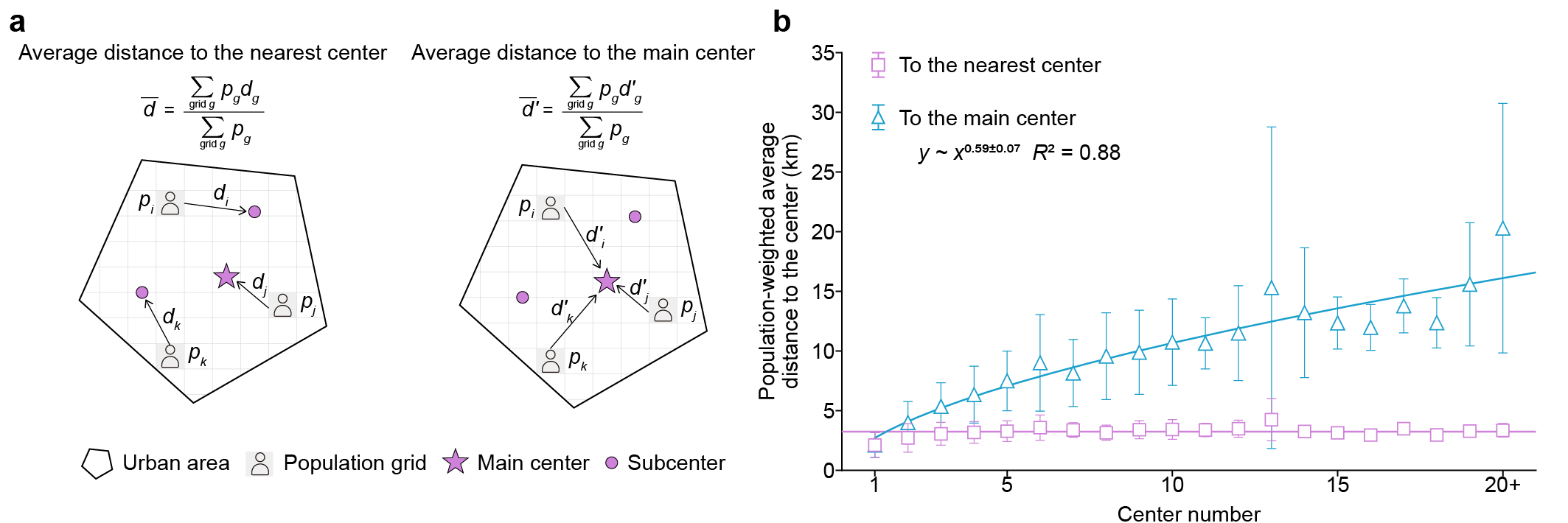}
    \captionsetup{font={footnotesize, stretch=0.8}}
    \caption{\textbf{Population-weighted average distance to the center.} (\textbf{a}) Two types of average distance. For the average distance to the nearest center, we overlay the population grid on the urban area, calculate the distance to the nearest center of each grid, and average it by the grid population. For the average distance to the main center, the distance to the main center of each grid is used. (\textbf{b}) Average distance relative to the center number of the urban area. Urban areas are grouped by the center number and those with at least 20 centers are grouped together. While the distance to the main center increases significantly with the number of centers, the distance to the nearest center remains stable. The scatters represent the mean and the error bars represent the standard deviation. The line for the main center represents the fitting to $y = ax^b$, and the line for the nearest center represents the average value of the means.}
\label{Distance}
\end{figure}
\clearpage

\subsection*{Acknowledgment}
We thank the support of the National Key Research and Development Program of China (Grant No. 2023YFB3906801), the National Natural Science Foundation of China (Grant No. 42422110), the Fundamental Research Funds for the Central Universities, Peking University. P.S. thanks the support by the Beijing Key Lab of Spatial Information Integration and Its Applications, Peking University.

\subsection*{Author contributions}

L.D. conceived the project; S.P. collected the data; S.P., J.Z., Y.L., and L.D. analyzed the data; L.D., S.P., and Y.L. wrote the manuscript.

\subsection*{Data availability}
The data used in this paper are publicly available; the nighttime light data can be downloaded from \url{https://eogdata.mines.edu/products/vnl/}, the urban area data can be downloaded from (\url{https://human-settlement.emergency.copernicus.eu/download.php}), the population data can be downloaded from Worldpop (\url{https://www.worldpop.org/}), the POI data can be downloaded from Foursquare OS (\url{https://opensource.foursquare.com/os-places/}), and the geonames data can be downloaded from GeoNames (\url{https://www.geonames.org/}). 

\subsection*{Code availability}
All analyses were performed in Python. The code developed for the analysis is available via GitHub at \url{https://github.com/urbansci/city-centers}.

\subsection*{Competing interests} The authors declare no competing interests.
\bibliographystyle{unsrt} 
\bibliography{ref.bib} 

\end{document}